\documentclass[pra,10pt,twocolumn,superscriptaddress,showpacs]{revtex4-1}
\usepackage{amsmath}
\usepackage{caption}
\usepackage{latexsym}
\usepackage{amssymb}
\usepackage{graphicx}
\usepackage[colorlinks=true, citecolor=blue, urlcolor=blue]{hyperref}
\usepackage{float}
\usepackage{amsfonts}

\newcommand{\ket}[1]{| #1 \rangle}

\begin{document}

\title{Necessary criterion for extracting thermodynamical work from qudit-entangled state}

\author{Sumit Nandi}
\email{sumit.enandi@gmail.com}
\affiliation{S. N. Bose National Centre for Basic Sciences, Block JD, Sector III, Salt Lake, Kolkata 700 106, India}
%
%

\begin{abstract}
 
 A novel criterion of extracting thermodynamical work from a bipartite pure qudit-entangled state by means of local operation and classical communication (LOCC) has been presented. We have shown that non-vanishing $G$-concurrence is a necessary condition to extract work from an higher dimensional entangled  state in LOCC paradigm. 
\end{abstract}


\maketitle

\section{Introduction}
Phenomenal inclusion of quantum information theory within the framework of thermodynamics opens up a scope to devise micro-engines \cite{scully} which can \textit{lawfully} draw work from a single heat bath. The important consequence of reading the microscopic degrees of freedom, in particular the orientation of spin of qubits, leads to extraction of work from a single heat bath \cite{szilard}. Needless to say, it does not violate second law of thermodynamics since entropy of the system increases as accumulated information is diminished with the extraction of work. However the given system may be classical or quantum by nature, but information originating out of quantum correlation can be declassified by suitable work extraction protocol which outperforms the system containing classical information \cite{zurek} only. Non-local aspects of quantum correlation is the main reason behind this, nonlocality in this case acts as a resource.  Information stemming out of non-locality is regarded as fuel as bits of information can be used to extract work from a correlated system attached to a single heat bath \cite{0jonathon}. Thus entanglement of the system can be exploited in performing particular thermodynamical work extracting protocol as it is done in information processing protocols. Some novel work extraction protocols from entangled states have been presented in \cite{jonathan}, \cite{funo}. In \cite{goold}, the authors proposed a protocol to extract work, known as ergotropy, from a quantum state with respect to some reference Hamiltonian, under cyclic unitaries, and shown that quantum correlation, in particular discord, in the system can enhance work that can be gained in the given paradigm. In the light of above discussion, the role of quantum correlation has been found effective for extracting maximum ergotropy. However, in our present work, we shall consider the novel work extraction protocol presented in \cite{Maruyama}. The author formulated  a suitable criterion of separability by deriving a Bell type equality  in terms of the maximum work that can be extracted from a bipartite state. Later this work had been carried forward for tripartite states \cite{1Maruyama}. 
We will generalise the protocol using qudits - higher dimensional quantum system. Notably, information processing protocols are often advantageous with qudits \cite{zhang}\cite{durt}. \\~\\
But the grain of truth is that the scope of qudit-entanglement is much broader - it is significantly different from qubit-entanglement quantitatively as well as qualitatively. For instance, any  qubit-entangled state satisfies famous positive partial transpose (PPT) criterion \cite{ppt}. Thus all entangled qubit states are negative partial transpose (NPT) which no longer remains sufficient for non-separability in qudit-entanglement \cite{ppt1}. There exists a class of states which are PPT and entangled as well, those are known as bound entangled states. Entanglement cannot be distilled from those states \cite{be} that raises a fundamental difficulties in detecting qudit-entanglement. In a similar manner, quantification of a single copy pure bipartite qubit-entanglement can be fully described by a single measure whereas a bipartite qudit state needs more such $ad$ $hoc$ quantities. The author \cite{vidal} had introduced a family of $d$ entanglement monotones to describe entanglement of a bipartite pure qudit state when only a single copy of the given state is provided. In a similar spirit the authors \cite{gour} had introduced concurrence monotones which are constructed with the Schmidt coefficients of the state. The last one, namely $G$-$concurrence$ had been a central attraction due to its extensive implication of full dimensionality of entanglement. It seems reasonable to think of the entangled states in a $d$ dimensional Hilbert space $C^d \otimes C^d$ which have $d^\prime(<d-1)$ non-vanishing entanglement monotones. So, some of the novel aspects of entanglement would be missing from those states. Meanwhile, it also implies that the performance of certain protocols, if carried out by such states, would reflect underlying structure of the state. To delve into more details, we have considered work extraction protocol by LOCC with bipartite qutrit ($d=3$) states. We have formulated a suitable criterion that ensures the success of the protocol. It is to be mentioned that we have shown for the first time that extractable work is directly related to a computational measure of entanglement, \textit{namely} G-concurrence. In doing so, we have also presented a physical interpretation of $G$-concurrence. We shall show entangled states with non-vanishing concurrence and $G$-concurrence are more suitable to extract work in non-local regime. 
\\
\\
Before presenting our main result, we shall revisit some prerequisites in Sec.(\ref{prerequisites}) which are very much relevant for our discussion. Then, we present the framework of a work extraction protocol by LOCC in Sec.(\ref{protocol}). Subsequently, we provide the main result in Sec.(\ref{theorem}) with some illustrative examples. Then we will make some important remarks in in Sec.( \ref{conclude}). 
\\
\\\section{Prerequisites}\label{prerequisites}
\subsection{Concurrence monotone}
Concurrence monotones \cite{gour} were introduced to characterise entanglement properties of a single copy pure entangled state of dimensionality beyond $2$. These are simply function of Schmidt coefficients of the given state. Thus one can realise all non-local aspects of a pure state by knowing all of these monotones. Concurrence monotones are computable and can be extended for mixed states by convex roof extension. For a pure state $\ket{\psi}$ in $C^d\otimes C^d$ having Schmidt numbers $\lambda_0$, $\dots$, $\lambda_{d-1}$ respectively, there exists $d-1$ non-trivial concurrence monotones. These monotones can be expressed in terms of a function of the Schmidt coefficients as
\begin{eqnarray}
C_1(\ket{\psi})&=&\sum_{i=0}^{d-1}\lambda_i\nonumber\\ C_2(\ket{\psi})&=&\sum_{i<j}^{d-1}\lambda_i\lambda_j \nonumber\\
.\nonumber\\
.\nonumber\\
C_d(\ket{\psi})&=&\prod_i^{d-1} \lambda_i.
\end{eqnarray} 
The first expression entails the fact that the first monotone is trivial and it is simply $one$, the sum of all Schmidt coefficients. Amongst the others, the last member of the monotone family is of particular interest and known as G-concurrence which can be recast as $C_d=d(\lambda_1\lambda_2..\lambda_d)^{\frac{1}{d}}$.
For $2\otimes 2$ pure states it is simply the concurrence as presented in \cite{Wooters}. For a $d$ dimensional state, G-concurrence is the determinant of the reduced density matrix of the subsystem. G-concurrence reveals the dimensionality of the entanglement. It may be possible that a $d$-dimensional state is entangled but its $C_d$ vanishes. By virtue of this fact, the state would be less effective for extracting information processing protocols.

\subsection{Thermodynamic work gain using entanglement as a resource }

In this subsection we elaborate the protocol \cite{Maruyama} to gain an insight about the role of entanglement in the context of thermodynamical work extraction by LOCC. Suppose, Alice and Bob shares a large number of bipartite states in distant lab scenario. Both of them have a number of measurement settings $\lbrace A_\theta, A^\perp_\theta\rbrace$  and $\lbrace B_{\theta^\prime}, B^\perp_{\theta^\prime}\rbrace$ respectively. Alice  measures her subsystem and let Bob know the outcome and measurement basis by classical communication. Knowing Alice's outcome, Bob can extract $1-H(B_{\theta^\prime}|A_\theta)$ bits of work from his subsystem. Here, $H(X|Y)$ is Shannon entropy of $X$ conditional on the entropy of $Y$. We can easily verify that  $1-H(B_{\theta^\prime}|A_\theta)$ is maximized when the state shared between Alice and Bob is maximally entangled $\ket\phi_{ME}$, $\ket\phi_{ME}=\frac{1}{\sqrt2}(\ket{00}+\ket{11})$. Quantitatively, the authors have shown the amount of extractable work from a given state $\rho$ is equivalent to
\begin{equation}
\zeta(A(\theta),B(\theta))=\frac{1}{2}(2-2H(A(\theta),B(\theta))+H(A(\theta))+H(B(\theta))),
\end{equation}
where $H(A(\theta))$ and $H(B(\theta))$ are the Shannon entropy corresponding to the outcome of $A(\theta)(B(\theta))$ respectively. Now one can Maximize $\zeta(A(\theta),B(\theta))$ by varying $\theta$ over the great circle of the Bloch sphere . Thus, extractable work is given by

\begin{equation}
\mathcal{W}(\rho)=\frac{1}{2\pi}\int_0^{2\pi}\zeta(A(\theta),B(\theta))d\theta.
\end{equation}
The quantity $\mathcal{W}$ carries a great deal of information about the non-locality of the state $\rho$. The maximal work $\mathcal{W}$ is bounded for separable states and entangled states can violate this upper bound. 
Now we extend this protocol for qutrit system \textit{i.e.} the shared state between Alice and Bob is a two-qutrit pure state. A measurement setting \cite{discord} can be constructed with the projectors $M_0$, $M_1$ and $M_2=\mathbb{I}-M_0-M_1$ where $M_i=|m_i\rangle\langle m_i|$ and
\\
\begin{eqnarray}
\ket{m_0}=e^{i\chi_1}\sin{\theta}\cos{\phi}\ket{0}+
e^{i\chi_2}\sin{\theta}\sin{\phi}\ket{1}+\cos{\theta}\ket{2},\nonumber\\
\ket{m_1}=e^{i\chi_1}\cos{\theta}\cos{\phi}\ket{0}+
e^{i\chi_2}\cos{\theta}\sin{\phi}\ket{1}-\sin{\theta}\ket{2},\nonumber
\end{eqnarray}   
\\ 
where $0\le\theta,\phi\le\frac{\pi}{2}$ and $0\le \chi_1,\chi_2\le2\pi$. For sake of simplicity, we have set $\chi_1,\chi_2=0$. It just takes a simple step to find the upper bound of work which is $(\simeq)$ $0.65$ for a separable state. Now we evaluate the quantity $\mathcal{W}$ for the following two-qutrit states.

\begin{eqnarray}
\ket{\tilde{\Omega}}&=&\sqrt{r}\ket{00}+\sqrt{s}\ket{11}+\sqrt{1-r-s}\ket{22} \\
\ket{\tilde{\omega}}&=&\sqrt{r}\ket{00}+\sqrt{s}\ket{11}+\sqrt{1-r-s}\ket{12},
\end{eqnarray}
where $r,s\in\{0,1\}$. Evidently $\ket{\tilde{\omega}}$ is entangled and its concurrence is given by $\sqrt{2(r-r^2)}$. Since its reduced density matrix has rank $2$, it is straightforward that $G$-concurrence vanishes for the given state. We calculate LOCC work that can be extracted from $\ket{\tilde{\omega}}$ for a given concurrence $0.9$ and it turns out to be  $\mathcal{W}(\ket{\tilde{\omega}})=0.50$. Although the state is highly entangled, as quantified by its concurrence, maximal work that can be gained from the state is no better than a separable one. Now for the former state we obtain $\mathcal{W}\simeq 0.8$ for given concurrence $0.9$. Therefore, merely specification of concurrence would not suffice to find the quantity $\mathcal{W}$ for a qudit state. It seems reasonable as entanglement property of higher dimensional quantum system is fully characterised by entanglement monotones.

\section{Thermodynamic work extraction protocol in LOCC paradigm}\label{protocol}
 Our approach of work extraction is pedagogical; the holistic protocol requires an entangled state being shared by two observers, namely Alice and Bob, and the former is performing a measurement in a pre-determined basis. Since, the shared state is entangled, the subsystem at Bob's end also changes  accordingly. Bob is provided with several filters which can transform the microscopic degree, such as orientation of spin of a qubit, into a specific orientation. As Bob is aware of the basis deployed by Alice, he picks up the suitable filter, in each turn Alice confirms her successful operation. Now, we consider Bob's subsystem after several such rounds of the protocol Fig.(a): after each turn of the process, Bob's subsystem resembles a specific state. So, Bob can extract work by allowing an isothermal expansion of his subsystem. If, on the other hand Bob cannot retrieve that specific state, we shall refer the work extraction protocol as an unsuccessful one. Let us describe the protocol more vividly with the maximally entangled two-qubit state $\ket{\phi}=\frac{1}{\sqrt2}(\ket{00}+\ket{11})$. Alice measures her subsystem in computational basis, as a result of it Bob's subsystem collapses into $\ket{0}$ and $\ket{1}$ corresponding to the measurement outcome $|0\rangle\langle0|$ and   
$|1\rangle\langle1|$ respectively. So, Bob does nothing for the former outcome while he applies $\sigma_x$ for the later outcome. Here the unitary operation $\sigma_x$ can be regarded as the filter chosen by Bob. So, in each turn Bob's state corresponds to a known state $\ket{0}$ which can be further exploited to extract thermodynamical work. In fact, this machinery would run for all measurement basis alike computational basis provided the shared state is maximally correlated. It is understood that the protocol does not work if a state like $\ket{00}$ is used instead, only half of the time Bob's subsystem resembles the pre-decided state. 
 
 We assume that Alice has performed operation on $N_0$ particles and each operation at Alice's end yields a particle in Bob's lab. So, at the end of the protocol, we might expect Bob's subsystem consists of $N_1$ $(\le N_0)$ objects with a specific orientation. The ratio $\frac{N_1}{N_0}$ would determine the amount of work to be extracted. Qualitatively it seems that the ratio would tend to $one$ when the resource state is maximally entangled and the ratio is less than unity for partially entangled states. Thus work extraction can be quantified from the entanglement of the resource state. We can consider $concurrence$ as a measure of entanglement for our present discussion.  
\\~\\
\begin{figure}\label{1figure}
\includegraphics[width=6.5cm,height=7cm]{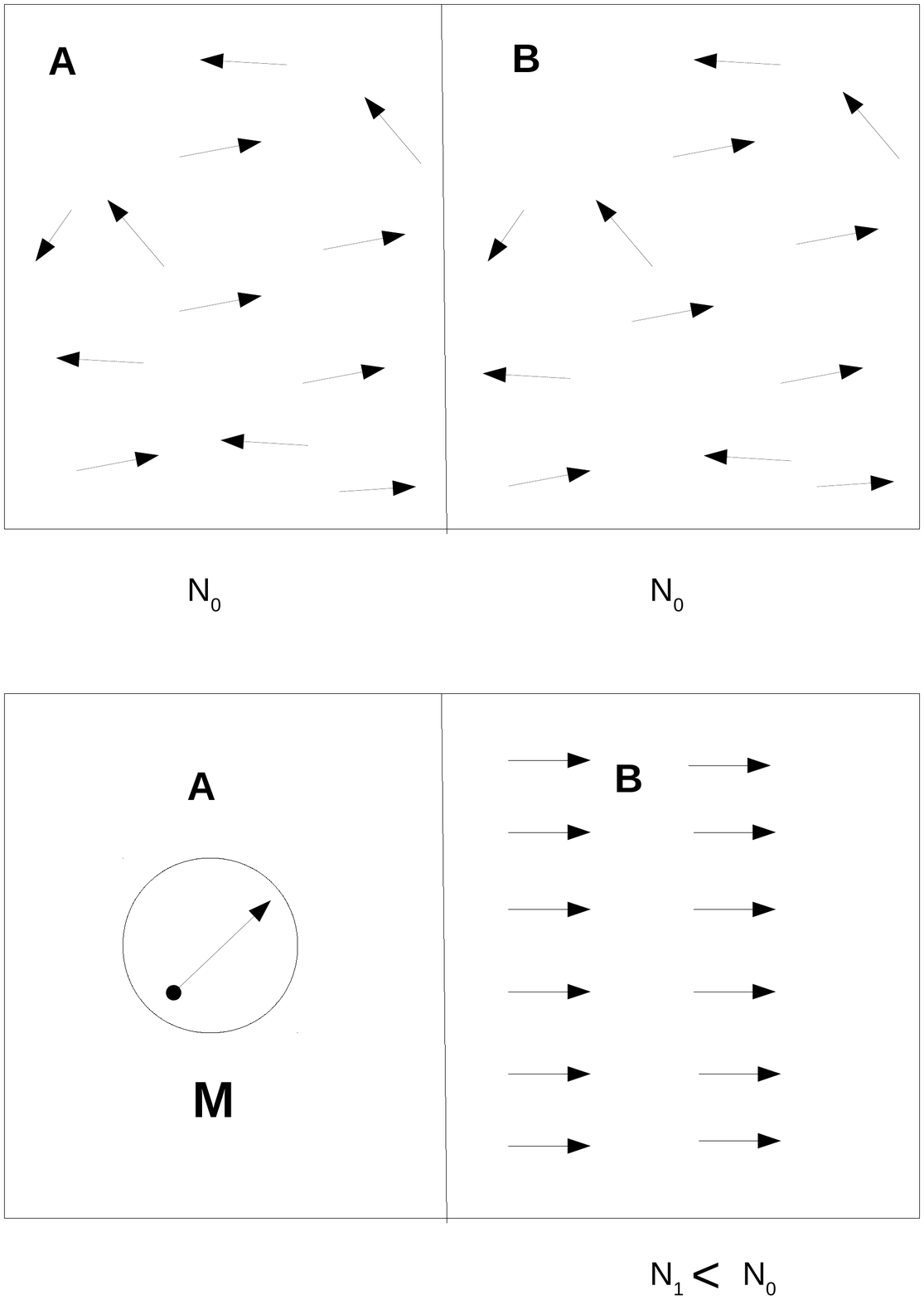}
\captionsetup{labelformat=empty}
\caption{(a)}
\end{figure}
In the above diagram, we have depicted the work extraction protocol: initially two observers are provided with two identical system composed of $N_0$ randomly polarised objects (represented by the arrowheads in the upper left and right boxes). Then the observer $\textbf{A}$ measures its subsystem and communicates $\textbf{B}$ the used measurement basis, according to which \textbf{B} applies a suitable filter and consequently its subsystem is filled with $N_1(\le N_0)$ objects with known polarisation state.\\~\\


   Now we substitute the working substance with the following qutrit entangled states given as
\begin{eqnarray}
\ket{\Omega}&=&\frac{1}{\sqrt{3}}(\ket{00}+\ket{11}+
\ket{22}),\label{st1}\\
\ket{\omega}&=&\frac{1}{\sqrt{3}}(\ket{00}+\ket{11}+
\ket{12})\label{st2} 
\end{eqnarray}
The state $\ket{\Omega}$ is maximally entangled two-qutrit state and  can be used to execute the protocol successfully. Following the preceding discussion, Alice measures her subsystem in computational basis and Bob applies identity, $\mathcal{O}_1$ and $\mathcal{O}_2$ corresponding to the outcome $|0\rangle\langle0|$, 
$|1\rangle\langle1|$ and $|2\rangle\langle2|$ respectively to retain his subsystem into the state $\ket{0}$. Here we explicitly write $\mathcal{O}_i$ as,
\begin{equation}
\mathcal{O}_1=
  \begin{bmatrix}
    0 & 1 & 0  \\
    0 & 0 & 1\\
    1 & 0 & 0
  \end{bmatrix} \hspace{.3in} \text{and} \hspace{.3in}
   \mathcal{O}_2=
  \begin{bmatrix}
    0 & 0 & 1  \\
    1 & 1 & 0\\
   0 & 1 & -1
  \end{bmatrix}
\end{equation}
  
Evidently the state $\ket{\omega}$ does not work under the given framework as Alice finds no output when she uses $|2\rangle\langle 2|$ to measure her subsystem. However, it would be possible to extract work if the role of Alice and Bob is reversed \textit{i.e.} Bob measures his subsystem and communicates with Alice. Nonetheless, less work can be gained as compared to the former state, since, Bob finds no such collapsed system in his lab to make measurement with whenever Alice's measurement outcome corresponds to that of $|2\rangle\langle 2|$. Henceforth, we shall define a successful work extraction protocol has occurred whenever Bob finds a definite subsystem collapsed in his lab corresponding to each measurement outcome of Alice. It is to be noted that $\ket{\omega}$ has non-vanishing concurrence, although is not a suitable resource for extracting thermodynamic work. Thus, it is an indicative  consequence of the fact that entanglement of a qutrit state quantified by concurrence is not sufficient to ensure the success of the protocol. In this spirit, we shall proceed to find a necessary criterion so that a qudit state can be used to extract work successfully in a thermodynamical system.

\section{Condition for maximal work extraction by LOCC}\label{theorem}
We proceed now to expose a criterion of extracting work in our paradigmatic situation. \\~\\
\textbf{Theorem:}
A pure bipartite qudit entangled state can be used 
as a resource state for extracting thermodynamic work by means of LOCC if it has non-vanishing G-concurrence.
\\
\\
To prove the above theorem, let us start with a general bipartite two-qudit state 
\begin{equation}\label{general_state}
\ket{\Psi}=\sum^{d-1}_{i,j=0}a_{ij}\ket{i}_A\ket{j}_B,
\end{equation}

where the complex coefficients $a_{ij}$ satisfy the normalization condition $\sum_{i,j}a^*_{ij}a_{ij}=1$. The subscripts $A(B)$ denote that the subsystems are distributed between Alice and Bob, respectively. The state written in the above form can always be expressed as follow
\begin{equation}
\ket{\Psi}=\sum^{d-1}_{\tilde{i}}\lambda_{\tilde{i}}
\ket{\tilde{i}}_A\ket{\tilde{i}}_B,
\end{equation}

where $\lambda_{\tilde{i}}$'s are the Schmidt coefficients of $\ket{\Psi}$. If Alice makes a measurement and communicates the outcome to Bob, then corresponding to each measurement outcome of Alice, the latter can find a suitable unitary $V$ to transform his subsystem into a specific state. While elaborating our work extraction protocol in the previous section, we found that a state would be useful for extracting work whenever, after application of a suitable unitary transformation, Bob's subspace corresponds to a definite state, say $|n\rangle\langle n|$, corresponding to Alice's each measurement outcome. This proves our theorem is sufficient for  accomplishing a successful work extraction protocol.\\~\\
Now, we will proceed to prove the necessity of the theorem.
As prescribed in the previous section, work extraction protocol begins with Alice who measures her subsystem with a general measurement setting consists of the projectors $m_0$, $m_1$, \dots , $m_{d}$ which satisfy $\sum^{d-1}_{i=0}|m_i\rangle\langle m_i|=\frac{\mathbb{I}}{d}$. Corresponding to each measurement outcome $i$ pertaining probability $p_i$ of Alice, Bob's subsystem is collapsed into the given one qudit state $\sum_j\tilde{a}_{ij}\ket{j}$, where $\tilde{a_{ij}}=\frac{a_{ij}}{\sqrt{p_i}}$. 
The protocol would be successful if Bob finds a suitable unitary $U$ such that the collapsed state is identified as a fixed reference state $\rho$ which can be expressed as $\rho=\sum_{n=0}^{d-1}|n\rangle\langle n|$ where $\sum_n|\langle n|n\rangle|=1$.  Here we emphasise that $n$ has same dimension as the index $j$.
 To accomplish this, we need a one-qutrit unitary operation $U$ such that
\begin{equation}\label{unitary}
\ket{j}=\sum_nU_{jn}\ket{n}.
\end{equation}
where $U_{jn}$ are the elements of $U$. The condition (\ref{unitary}) points towards an interesting properties of the complex elements $a_{ij}$. It would be possible if $a_{ij}$ satisfy the following condition 
\begin{equation}\label{condition}
\sum_ia^*_{i j^\prime}a_{ij}=\delta_{j
j^\prime},
\end{equation}
 To check this, let us consider Bob's subsystem $\rho_B$ by tracing over Alice's one as
 \begin{equation}
\rho_B =\sum_{i,j,j^\prime}a^*_{i j^\prime}a_{ij}|j\rangle\langle j^\prime|
 \end{equation}
 Now invoking Eqs. $(\ref{unitary})$ and $(\ref{condition})$ we find 
\begin{eqnarray}\label{Bob_state_3}
\rho_B &=& \sum_{j,j^\prime,n,n^\prime}\delta_{j
j^\prime}U^*_{j^\prime n^\prime}U_{jn}\ket{n}\langle {n^\prime}|\\
&=& \sum_{j,n,n^\prime}U^*_{jn^\prime}
U_{jn}\ket{n}\langle {n^\prime}|.
\end{eqnarray}
We put the condition of unitary $\sum_{j}U^*_{jn^\prime}U_{jn}
=\delta_{nn^\prime}$ and plug it into the last equation to obtain
\begin{eqnarray}
\rho_B &=& \sum_{n,n^\prime}\delta_{nn^\prime}\ket{n}\langle {n^\prime}|\\
&=& \sum_n \ket{n}\langle {n}|,\label{eq_10}\\
&=&\rho.
\end{eqnarray}
Thus we obtain a suitable criterion of successful work extracting protocol of the complex coefficients $a_{ij}$ of the state ({\ref{general_state}}).
The condition implies that Bob's subspace constitutes an orthonormal basis. Now it suffices to evaluate G-concurrence of the given state ({\ref{general_state}}).
We know that $G$-concurrence of a state of the form (\ref{general_state}) can be expressed as $G\varpropto \prod_{j=0}^{d-1}|\langle j|j\rangle|$. Thus invoking the condition (\ref{condition}), we obtain the resource state has non-vanishing G-condition which turns out to be necessary to carry out the prescribed task with the state (\ref{general_state}). This completes the proof of the theorem. \\~\\
   
The above theorem implies that the given state would be useful for the task if all of its Schmidt coefficients exist. For two level quantum system it would simply imply that $\ket{\Psi}$ is entangled with non-vanishing concurrence. Beyond two level system concurrence seems to be insufficient to ensure work extraction protocol. The state with vanishing G-concurrence would not be a potential resource state to extract thermodynamical work in the given paradigm as it was shown in the last section: the state $\ket{\Omega}$ was found to be useful for the protocol whether $\ket{\omega}$ was not as suitable.\\~\\ 

To assert our theorem, we shall extend the protocol \cite{Maruyama} 
by two well known classes of mixed states to show the quantity $\mathcal{W}$ is a monotone of G-concurrence.
\\
\\
\textit{Example:} Here we consider the state $\rho=x|\omega\rangle\langle\omega|+(1-x)|\Omega\rangle\langle\Omega|$.  We mention that G-concurrence of $\rho$, for this particular decomposition, can be given as $1-x$. A quantitative behaviour of the quantity $\mathcal{W}(\rho)$ for the given state is depicted in Fig.(b). The plot shows that $\mathcal{W}(\rho)$ in non-local regime decreases as $x$ increases. Here $x$ weights the presence of the state $\ket{\omega}$ which has vanishing G-concurrence.  \\~\\

\begin{figure}\label{2figure}
\includegraphics[width=6.5cm,height=5cm]{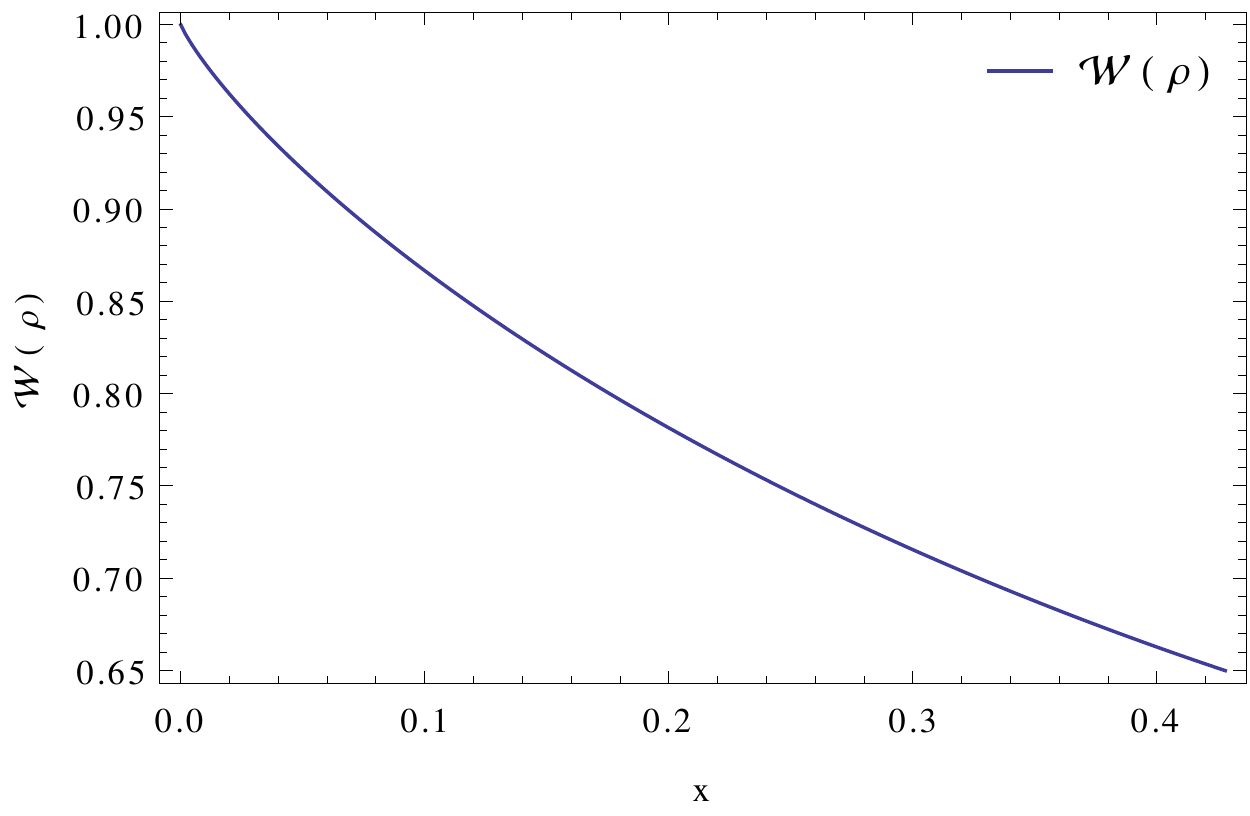}
\captionsetup{labelformat=empty}
\caption{(b)}
\end{figure}

\textit{Example:} Lastly we shall produce an example to show our criterion (\ref{condition}) is robust against white noise, we consider the mixed state 
\begin{equation}
\rho_{mix}=\alpha|\Omega\rangle\langle\Omega|+(1-\alpha)|01\rangle\langle01|. 
\end{equation}
Although there does not exist any general prescription to compute concurrence monotones of arbitrary mixed states, the G-concurrence of the given state  $\rho_{mix}$ has been found in \cite{gour1} and it is given by $\alpha$.  We plot the quantity $\mathcal{W}(\rho_{mix})$ for the given state $\rho_{mix}$ and show in the Fig.(c). It follows that extractable work $\mathcal{W}(\rho_{mix})$ monotonically increases with its G-concurrence. 
\\~\\

\begin{figure}\label{3figure}
\includegraphics[width=6.5cm,height=5cm]{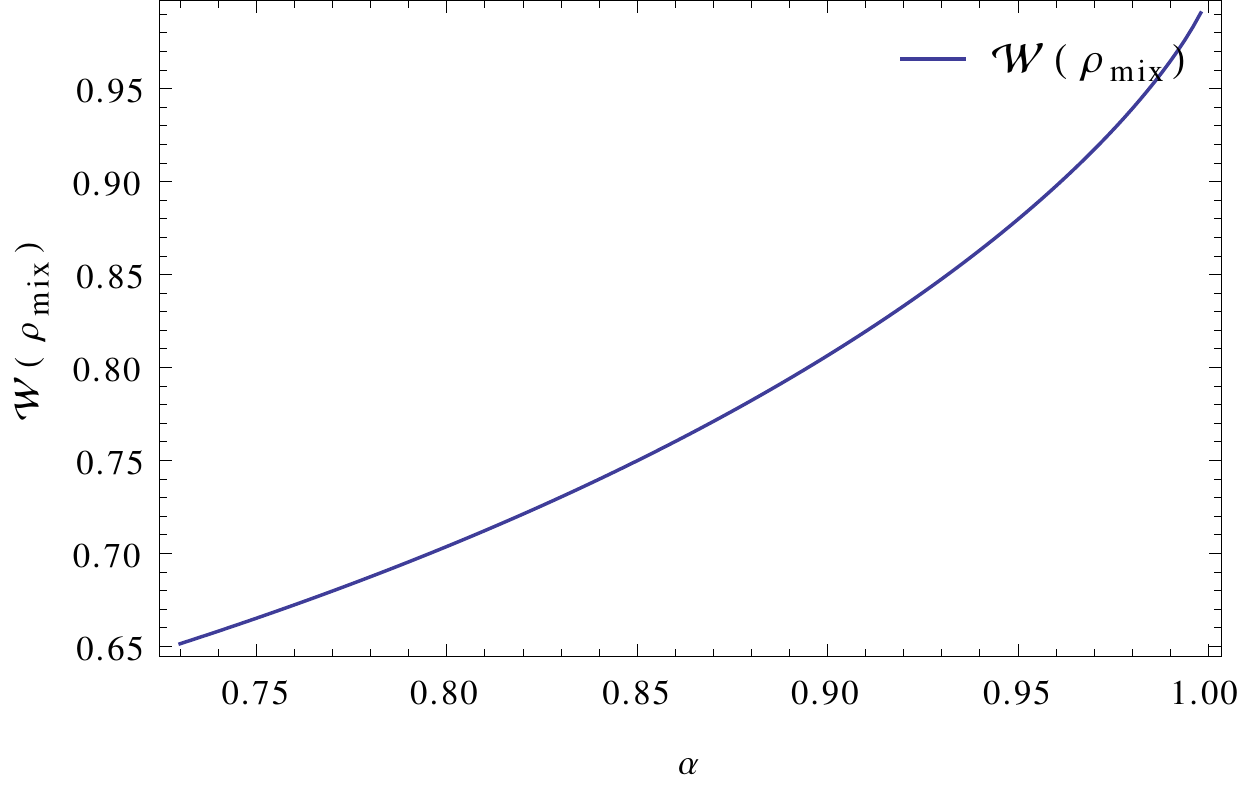}
\captionsetup{labelformat=empty}
\caption{(c)}
\end{figure}


\section{Conclusion}\label{conclude}
We have found a necessary criterion to extract thermodynamical work form a qutrit entangled state. It turns out that non-vanishing G-concurrence is necessary for a state so that it can be used as a resource for the prescribed protocols. It is to be noted that the result can also be generalised straightforwardly for any qudit states. The result is not so surprising in itself as we know that specification of a family of entanglement monotones is necessary to characterise the entanglement property of a qudit state. So a state with high concurrence value but vanishing G-concurrence is not suitable for extracting work in thermodynamic regime. We have explicitly found the necessary condition to extract work successfully. Apart from a proof in full generality, suitable examples have been constructed to strengthen our result. We have  provided few classes of mixed states to show that the quantity $\mathcal{W}$ monotonically increases with G-concurrence. Thus we have also obtained a physical interpretation of G-concurrence of qudit entangled states. We mention that extensive characterisation of entanglement in qudit regime needs G-concurrence, perhaps other monotones also which may occur beyond qutrit states. In our discussion, we have seen that entanglement might be present in the subspaces of a qutrit state which is quantified by its concurrence. Indeed it appears to be insufficient to accomplish certain LOCC protocol 
while entanglement arising out of full dimensionality of the state space is very much required.     
\section*{Acknowledgements}The author acknowledges support from the Department of Science
and Technology, Government of India through the QuEST
grant (project no.-
DST/ICPS/QuEST/2018/98).

\end{document}